\documentstyle[sprocl]{article}

\input{psfig}

\def\Journal#1#2#3#4{{#1} {\bf #2}, #3 (#4)}

\def\NPB{{\em Nucl. Phys.} B}
\def\PLB{{\em Phys. Lett.}  B}
\def\PRL{\em Phys. Rev. Lett.}
\def\PRD{{\em Phys. Rev.} D}

\def\be{\begin{equation}}
\def\ee{\end{equation}}
\def\bea{\begin{eqnarray}}
\def\eea{\end{eqnarray}}

\begin{document}

\title{Laplacian Center Vortices\footnote{presented by Ph. de Forcrand.} }

\author{Ph. de Forcrand}

\address{Inst. f\"ur Theoretische Physik, ETH H\"onggerberg, 
CH-8093 Z\"urich, Switzerland \\
and \\
CERN, Theory Division, CH-1211 Gen\`eve 23, Switzerland \\
E-mail: forcrand@itp.phys.ethz.ch} 

\author{M. Pepe}

\address{Inst. f\"ur Theoretische Physik, ETH H\"onggerberg, 
CH-8093 Z\"urich, Switzerland \\
E-mail: pepe@itp.phys.ethz.ch} 


\maketitle\abstracts{
I present a unified picture of center vortices and Abelian monopoles.
Both appear as local gauge ambiguities in the Laplacian Center Gauge.
This gauge is constructed for a general $SU(N)$ theory. Numerical
evidence is presented, for $SU(2)$ and $SU(3)$, that the projected $Z_N$ 
theory confines with a string tension similar to the non-Abelian one.
}

\section{Motivation and technical problem}

The road traveled by physicists in their efforts to identify the
effective, InfraRed degrees of freedom of QCD is far from straight.
It bifurcates in many directions, most of which are still under
exploration. Currently, the two most popular effective descriptions
of confinement are in terms of Abelian monopoles \cite{tH_Ab,Mandelstam} and
center vortices \cite{tH_CV,Mack}. I want to show that these two descriptions
can be unified: these two branches of the road merge together, perhaps
indicating that we are traveling towards a piece of Truth.

The study of center vortices, first proposed by Mack \cite{Mack} and
by 't Hooft \cite{tH_CV}, has been revived by Greensite and collaborators.
They project the $SU(2)$ lattice gauge theory to a $Z_2$ gauge theory,
by partially fixing the gauge to Maximal Center Gauge, defined as the
gauge in which
\be
\sum_{x,\mu} |Tr~U_\mu(x)|^2 ~~{\rm maximum} \quad .
\label{MCG}
\ee
In this gauge, the center-projected links are 
$z_\mu(x) \equiv {\rm sign}(Tr~U_\mu(x))$.
This $Z_2$ theory has defects corresponding to plaquettes taking value $-1$.
Greensite et al.\cite{Green} showed that the string tension $\sigma$ given 
by these defects 
closely matches that of the original $SU(2)$ theory. Being skeptical about
this, I investigated with M. D'Elia the coset theory, made of positive-trace
links $U'_\mu(x) \equiv z_\mu(x) ~ U_\mu(x)$. This theory has more 
short-range disorder than the original one, but carries no center vortices.
Could it be that $\{U'_\mu(x)\}$ would show long-range order, and thus 
not confine? To my surprise, we found \cite{PRL} that in $\{U'_\mu(x)\}$:
$(i)$ $\sigma = 0$; $(ii)$ $\langle \bar\psi \psi \rangle = 0$;
$(iii)$ $Q_{top} = 0$. Removal of center vortices causes
deconfinement, chiral symmetry restoration and suppression of topological
excitations. All non-perturbative properties disappear. Therefore, 
center vortices must carry the non-perturbative degrees of freedom.

There are two difficulties with this conclusion. First, a great deal of
numerical evidence has been accumulated which ties confinement with Abelian
monopoles instead of center vortices. Secondly, the above findings may depend
on the choice of local maximum in (\ref{MCG}). It is the purpose of this 
talk to resolve these two difficulties, as outlined already in \cite{Pisa}.

The second problem is shared by the Abelian projection, which also proceeds
by gauge fixing via the iterative, local maximization of a gauge functional.
But in the Maximal Center Gauge, this problem can be acute. As shown in
\cite{KT}, the local maximum reached after starting from Landau gauge leads 
to a very small density of center vortices, which actually do not confine.
Following this severe warning, some studies try to obtain the center-vortex  
properties of the global maximum of (\ref{MCG}) by taking the highest among
$m$ local maxima, and extrapolating to $m \rightarrow \infty$ \cite{Bornyakov}.
One feature underlines the difficulty of this approach: the extrapolated value
for the global maximum of (\ref{MCG}) falls {\em below} the measured value
obtained by the procedure of ~\cite{KT}.

To illustrate why this technical problem is so hard to resolve, let us consider 
a toy example. Take a 1-dimensional ring of $U(1)$ links 
$\{e^{i\theta_i},i=1,..,N\}$ such that the gauge invariant loop which they
form is $-1$: $\prod_N e^{i\theta_i} = -1$. For such a system, the global
maximum of (\ref{MCG}), corresponding to Maximal Center Gauge, is obtained
when $\theta_{i_0} = \pi$ for one link $i_0$, and 
$\theta_i = 0 ~\forall i \neq i_0$. The ``kink'' $\theta_{i_0} = \pi$ can be placed
anywhere, giving rise to an $N-$fold degeneracy. The gauge-fixing functional
(\ref{MCG}) takes value $N$. Let us now fix this system to Landau gauge,
defined as the gauge which maximizes $\sum_N ReTr~U_i$. This is achieved 
when $\theta_i = \pi/N$. All link angles are small and there is no sign of
a kink. The Center Gauge functional (\ref{MCG}) then takes the value 
\be
\sum_N |Tr~U_i|^2 = N cos\frac{2\pi}{N} \approx N - \frac{2\pi^2}{N}
\label{toy}
\ee
One can check that Landau gauge represents a local maximum of (\ref{MCG});
we see that the difference between this local maximum and the global one is 
vanishingly small as $N \rightarrow \infty$.

This case is not just a toy example. An Abelian loop having a phase of $\pi$
is precisely the signature of an $SU(2)$ center vortex, as illustrated in Fig.1.
It is therefore essential to address the technical problem of gauge-fixing
in order to properly identify the center vortices.

\begin{figure}[t]
\begin{center}
\epsfig{figure=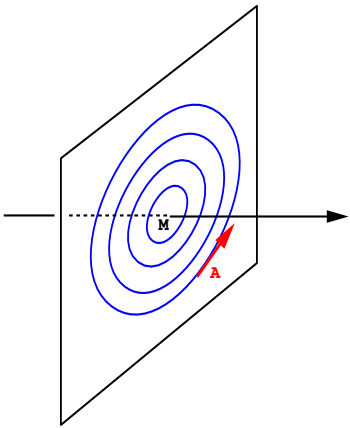,height=2.3in,width=1.9in}
\end{center}
\caption{A ``prototype'' $SU(2)$ center vortex: the tangential 
gauge potential is locally Abelian, $A_\phi^3(r) = \frac{1}{2r}$, 
so that an $SU(2)$ Wilson loop pierced by the vortex acquires
a minus sign.
\label{protoCV}}
\end{figure}

\section{Unambiguous gauge-fixing}

The crucial observation is that, since a center vortex gives the fundamental
Wilson loop a factor $-1 = {\rm exp}(i \pi \sigma_3)$, it will give the
adjoint Wilson loop a factor ${\rm exp}(2 i \pi \dot{\sigma_3})$. Therefore, as
one describes a small loop around point $x_0$ in Fig.1, the adjoint gauge
field winds by $2 \pi$ in color space. Thus, the center vortex at $x_0$ will appear
as a gauge singularity of the adjoint field. More formally, one can observe
that center vortices correspond to non-trivial topological classes 
of $\Pi_1(SU(N)/Z_N)$,
so that Wilson loops in the ``center-blind'' adjoint representation 
are the proper objects to identify them.

A second observation motivates our gauge-fixing approach. If the gauge field
$A_\mu^{adj}$ diverges like $\frac{1}{r}\dot{\sigma_3}$ near $x_0$, then so will
the covariant derivative $D_\mu^{adj} = \partial_\mu + i g A_\mu^{adj}$. This will
force the vector $(D_\mu^{adj} \vec{v})(x)$ to orient itself along
$\dot{\sigma_3}$ when $x$ approaches $x_0$, for any smoothly varying vector 
$\vec{v}(x)$. In particular, this statement applies to all the eigenvectors
of any adjoint covariant derivative operator. In the vicinity of a thin center 
vortex similar to the ``prototype'' of Fig.1, these
eigenvectors will become {\em collimated} in color space.
In reality, the singularity is smoothed out: the center vortex is no longer
``thin'', but acquires a finite size. Collinearity of eigenvectors still
takes place within the vortex core, but at an exact location which depends
on the particular choice of covariant operator and eigenvectors.

This collinearity property can be used to detect center vortices on the lattice,
without actually performing any gauge fixing or center projection. For
simplicity, we take as covariant operator the adjoint Laplacian, discretized
in the simplest way:
\be
\Delta_{xy}^{adj} = \sum_{\pm\mu} \dot{U}_{\pm\mu}(x) \delta_{y,x\pm\mu}
                   - 2 d \delta_{y,x} \quad ,
\ee
where $\dot{U}_{ab} = \frac{1}{2} Tr( \tau_a U \tau_b U^\dagger)$ is the link
in the adjoint representation, and the $\tau$'s are the generators of $SU(N)$.
The signal for an $SU(2)$ center vortex is the parallelism in color space of two
eigenvectors.
Of course, this will never occur exactly at a lattice site, and interpolation
is necessary. To reduce interpolation ambiguities, we choose the two lowest-lying
eigenvectors, which are the least sensitive to UV fluctuations.

Because we want to study the projected $Z_N$ theory, we proceed with gauge fixing.
We can also make use of the eigenvectors of the Laplacian for that purpose
instead of maximizing (\ref{MCG}).
Since the Laplacian is covariant, a local gauge transformation $\Omega(x)$,
which transforms the gauge links $U_\mu(x)$ into 
$\Omega(x) U_\mu(x) \Omega^\dagger(x+\hat\mu)$,
will also transform the eigenvector $v(x)$ into $\Omega(x) v(x)$. Following Ref.\cite{VW},
we can then fix the gauge {\em uniquely} by specifying the orientation of $\Omega(x) v(x)$
in color space at each point $x$. This kind of gauge was called Laplacian gauge
in \cite{VW}. We generalize it below to the adjoint representation. When the adjoint
$SU(N)/Z_N$ field is fully gauge-fixed, a remnant center gauge freedom $Z_N$ 
subsists, and we can look at the projected $Z_N$ gauge theory.

Gauge fixing proceeds in two steps. \\
{\bf Step 1}. At each point $x$, the lowest-lying eigenvector $v^{(1)}(x)$ has
$(N^2-1)$ real color components. Let us form the hermitian matrix
\be
\Phi^{(1)} \equiv \sum_a^{N^2-1} v^{(1)}_a \tau_a
\ee
and apply the gauge transformation $\Omega(x)$ which diagonalizes $\Phi^{(1)}$
at every $x$. After this step, $v^{(1)}$ will be rotated to lie along the diagonal
generators of $SU(N)$: it will be parallel to $\sigma_3$ in the case of $SU(2)$,
or lie in the plane $(\tau_3,\tau_8)$ for $SU(3)$. This gauge-fixing is not complete:
in the $SU(2)$ case, an arbitrary rotation $e^{i \theta \sigma_3}$ can still be
performed. More generally, after one fixes some arbitrary ordering for the
eigenvalues of $\Phi^{(1)}$, there remains an Abelian $U(1)^{N-1}$ gauge freedom.
In other words, we have reduced the gauge symmetry to its maximal Cartan subgroup.
This gauge was proposed for $SU(2)$ by A. van der Sijs, who called it Laplacian Abelian Gauge,
as an unambiguous substitute for Maximal Abelian Gauge \cite{AvdS}: it achieves 
the same purpose but has no difficulty with gauge copies coming from local maxima. 
Here it appears as a natural intermediate step, which exhausts the information
available from the lowest-lying eigenvector.

{\bf Step 2}. Using the next eigenvector rotated by $\Omega(x)$, 
$\tilde{v}^{(2)}(x) = \dot{\Omega}(x) v^{(2)}(x)$, we form the matrix
\be
\Phi^{(2)} \equiv \sum_a^{N^2-1} \tilde{v}^{(2)}_a \tau_a \quad .
\ee
An Abelian rotation will leave the diagonal part of $\Phi^{(2)}$ invariant.
Therefore, we complete the gauge fixing by enforcing $(N-1)$ constraints on 
off-diagonal elements, which we choose to be on the sub-diagonal. Specifically, 
for $SU(2)$ we require that $\tilde{v}^{(2)}$
lie in the positive $(\sigma_1,\sigma_3)$ half-plane: 
$Tr \Phi^{(2)} \sigma_2 = 0,~Tr \Phi^{(2)} \sigma_1 > 0.$
Similarly, for $SU(3)$ we require
$Tr \Phi^{(2)} \tau_2 = Tr \Phi^{(2)} \tau_7 = 0,~Tr \Phi^{(2)} \tau_1 > 0,~
Tr \Phi^{(2)} \tau_6 > 0$.
This 2-step procedure fixes the gauge completely, up to local $Z_N$ transformations
which leave the adjoint links and the Laplacian eigenvectors unchanged.

\section{Local gauge ambiguities}

Laplacian Center Gauge is unambiguous: no matter what the starting point on the gauge
orbit, the gauge-fixed configuration will be the same (up to a global gauge
transformation). For some exceptional gauge fields, the smallest eigenvalues
$\lambda_1, \lambda_2$ may be degenerate, leading to a continuous family of
possible gauge-fixed solutions, which can legitimately be called Gribov copies.
But this situation never occurs in practice.

What occurs for a generic gauge field, however, is that the gauge transformation
becomes ill-defined at some point(s) $x$. These {\em local} gauge ambiguities
prevent a complete gauge fixing at $x$; they locally enlarge the gauge symmetry
beyond $Z_N$. As we shall now see, these gauge defects are center vortices and monopoles.

One of the constraints of Step 2 is automatically satisfied if the associated complex 
matrix element is zero.
In that case the Abelian gauge rotation cannot be
fixed, and the remaining gauge symmetry is enlarged from $Z_N$ to $U(1)$. Specifically,
for $SU(2)$ this occurs when $\tilde{v}^{(2)}_1 = \tilde{v}^{(2)}_2 = 0$, so that
$\tilde{v}^{(2)}$ lies along the color direction $\sigma_3$ just like $v^{(1)}$.
A $U(1)$ gauge ambiguity occurs whenever $v^{(1)}$ and $v^{(2)}$ are collinear.
One can check that a small Wilson loop around such a point will acquire a phase
$e^{i 2 \pi \dot{\sigma_3}}$ in the adjoint representation, or $e^{i \pi \sigma_3} = -1$
in the fundamental. We have recovered the statement, made at the beginning of Section 2,
that center vortices can be detected by the collinearity of eigenvectors. Collinearity
implies 
$\frac{v^{(2)}_1}{v^{(1)}_1} = \frac{v^{(2)}_2}{v^{(1)}_2} = \frac{v^{(2)}_3}{v^{(1)}_3}$,
so that 2 constraints must be satisfied. Center vortices have codimension 2, forming
closed surfaces in $4d$. In the case of $SU(3)$, ambiguities happen when
$\tilde{v}^{(2)}_1 = \tilde{v}^{(2)}_2 = 0$, or $\tilde{v}^{(2)}_6 = \tilde{v}^{(2)}_7 = 0$. 

\begin{figure}[t]
\begin{center}
\epsfig{figure=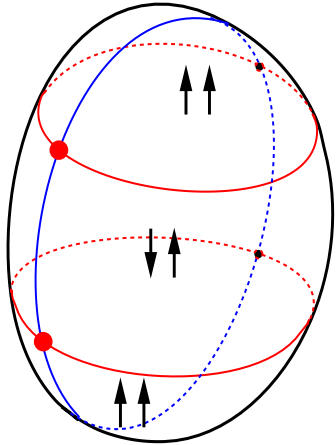,height=2.3in,width=1.7in}
\end{center}
\caption{The connection between center vortices and Abelian monopoles:
the (horizontal) monopole worldline separate two patches of center vortex 
surface with
opposite eigenvector orientations; each monopole is attached to two
center vortex strings.}
\end{figure}

Local ambiguities can already arise at Step 1. For $SU(2)$, this happens whenever
$|v^{(1)}(x)| = 0$, yielding 3 constraints. These defects have codimension 3, forming
closed loops in $4d$. Along such lines the Abelian projection is ill-defined.
't Hooft \cite{tH_Ab}, followed by \cite{AvdS}, has shown that these defects can be
identified with monopole worldlines. In the case of $SU(3)$, the analysis is more 
subtle\cite{progress}. Whenever two eigenvalues of $\Phi^{(1)}$ become degenerate, 
their ordering
becomes ambiguous. This enlarges the remaining gauge freedom from $U(1)^{N-1}$ to
$U(1)^{N-2}\times SU(2)$. Again, these defects can be identified with codimension-3
monopole worldlines.

Now an intimate connection appears between monopoles and center vortices. 
The latter correspond to $v^{(1)}$ and $v^{(2)}$ being collinear. The two vectors
can be parallel or anti-parallel. To go from one situation to the other along the
center vortex surface, one must cross a line where $|v^{(1)}|=0$ or $|v^{(2)}|=0$.
Thus, monopole worldlines always separate patches of center vortices with oppositely
oriented eigenvectors. This situation is schematically described in Fig.2.
One can interpret center vortices as the worldsheet of half Dirac strings between
two monopoles. This explains the numerical finding that almost all monopoles
are attached to two center vortices \cite{Stack}.

\section{Numerical findings}

We have applied the gauge-fixing procedure described above to $SU(2)$ and $SU(3)$
ensembles. The projected $Z_2$ or $Z_3$ ensembles are constructed by replacing each
link by the center element whose trace is the closest. To estimate the string 
tension in the projected theory, we measure the Creutz ratios
$\chi_{R,R} \equiv - \ln [\langle W_{R,R} \rangle \langle W_{R+1,R+1} \rangle
/\langle W_{R,R+1} \rangle^2]$, where $\langle W_{R,T} \rangle$ is the 
expectation value of an $R$ by $T$ Wilson loop.

\begin{figure}[t]
\begin{center}
\epsfig{figure=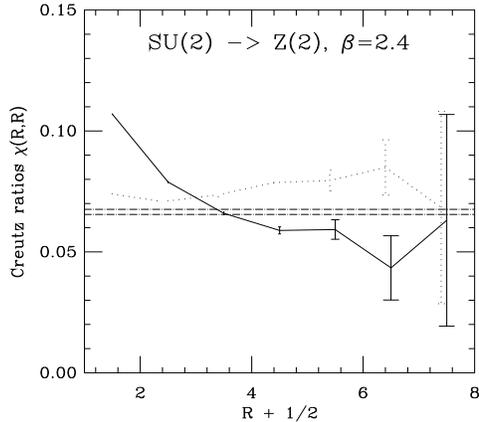,height=2.0in,width=2.5in}
\end{center}
\caption{$Z_2$ Creutz ratios via Direct Maximal Center gauge (dotted line)
and Laplacian Center gauge (solid line) on a $16^4$ lattice. The dotted band 
shows the $SU(2)$ string tension.}
\end{figure}
\begin{figure}[h]
\begin{center}
\epsfig{figure=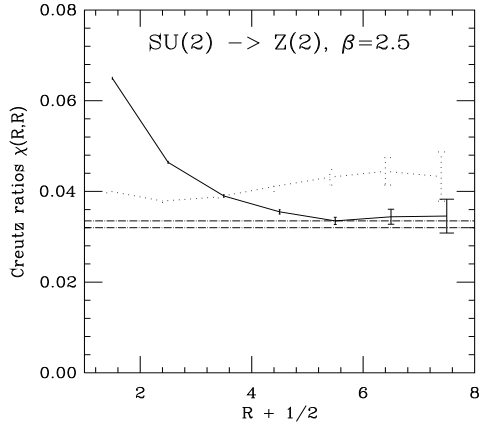,height=2.0in,width=2.5in}
\end{center}
\caption{Same as Fig.3, $\beta=2.5$.}
\end{figure}

We first performed a check of the usual Direct Maximal Center gauge approach\cite{DMC}.
As illustrated in Figs.3 and 4, the Creutz ratios in the projected $Z_2$ theory appear 
to {\em grow} with distance. This trend can already be noticed in Fig.2 of Ref.\cite{PRL}.
It has also been observed in the projected $U(1)$ theory obtained after Maximal
Abelian gauge-fixing \cite{AvdS}. It may depend on the details of the iterative
gauge-fixing algorithm. In any case, increasing Creutz ratios signal a lack of 
positivity of the transfer matrix. Positivity is of course not guaranteed after
a non-local procedure such as center gauge-fixing and projection. We mention this
as a potential problem of the iterative gauge-fixing approach, in addition to that
of local maxima. As a final ``nail in the coffin'', we should add that Laplacian
Center gauge fixing is cheaper: on a $16^4$ lattice, using the package 
ARPACK\cite{arpack} to find the necessary eigenvectors, the work performed to fix the
gauge is equivalent to about 50 Monte Carlo sweeps
for $SU(2)$, and 300 to 500 for $SU(3)$; this is far less than required by the
iterative approach.

\begin{figure}[h]
\begin{center}
\epsfig{figure=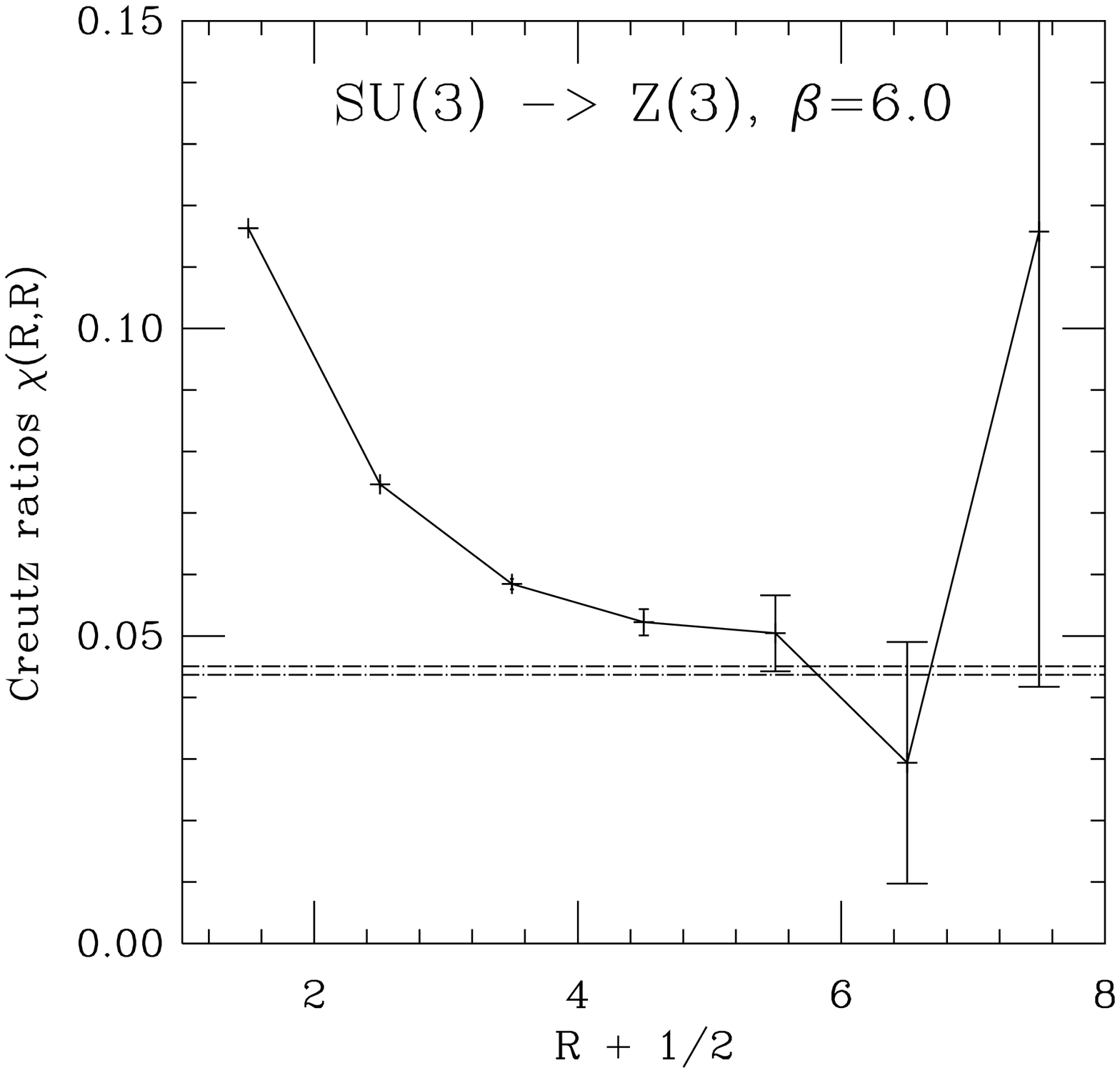,height=1.98in,width=2.5in}
\end{center}
\caption{Same as Fig.3, here for $SU(3), \beta=6.0$.}
\caption{$Z_3$ Creutz ratios via Laplacian Center gauge on a $16^4$ lattice. 
The dotted band shows the $SU(3)$ string tension.}
\end{figure}

Figs.3 to 5 show the Creutz ratios in the projected theory after Laplacian Center
gauge fixing, for $SU(2)$ ($\beta=2.4$ and $2.5$) and $SU(3)$ ($\beta=6$).
As the distance increases, the Creutz ratios decrease rapidly then stabilize.
The rapid decrease indicates the presence of many close pairs of vortices, quite
unlike the Direct Maximum Center gauge results. Indeed, the vortex density is about
a factor of 2 greater. This is in line with the increased monopole density observed
in Laplacian Abelian gauge\cite{AvdS}. The plateau reached by the Creutz ratios
appears close to the full $SU(2)$ or $SU(3)$ string tension. A more extensive 
numerical study is under completion\cite{progress}.

The significance of the rough agreement between the $Z_N$ and $SU(N)$ string tensions
should not be exaggerated. Our measurements are taken at some finite,
rather large value of the lattice spacing. We could choose another gauge,
defined for example via a different discretization of the Laplacian, with
higher derivative terms. We should expect the $Z_N$ Creutz ratios to be 
sensitive to this choice of gauge. It is only in the continuum limit that we
must recover gauge independence and agreement between the $Z_N$ and $SU(N)$ 
string tensions.

\section*{Acknowledgments}
We thank C. Alexandrou, S. D\"urr, M. D'Elia and J. Fr\"ohlich for
helpful discussions.

\end{document}